\documentclass[twocolumn,showpacs,preprintnumbers,amsmath,amssymb,pra]{revtex4}

\usepackage{graphicx}
\usepackage{dcolumn}
\usepackage{bm}
\usepackage{mathrsfs}
\usepackage{amssymb}
\usepackage{bbm}
\usepackage{pstricks}

\newcommand{\bbP}{\hat{\mathbb{P}}}
\newcommand{\bbQ}{\hat{\mathbb{Q}}}
\renewcommand{\L}{\hat{\mathscr{L}}}
\renewcommand{\H}{\hat{H}}
\newcommand{\R}{\hat{R}}

\renewcommand{\a}{\hat{a}}
\newcommand{\oTheta}{\hat{\Theta}}
\newcommand{\osigma}{\hat{\sigma}}
\newcommand{\orho}{\hat{\rho}}
\newcommand{\eye}{\hat{\mathbbm{1}}}

\newcommand{\Tr}{{\rm Tr}}

\newcommand{\ket}[1]{|#1\rangle}
\newcommand{\bra}[1]{\langle #1|}

\begin{document}

\preprint{APS/123-QED}

\title{Open quantum systems approach to atomtronics}

\author{R. A. Pepino, J. Cooper, D. Meiser, D. Z. Anderson, and M. J.
Holland}

\affiliation{JILA, National Institute of Standards and Technology and
  Department of Physics, The University of Colorado, Boulder, Colorado
  80309-0440, USA}

\date{\today}

\begin{abstract} We derive a quantum master equation to treat quantum systems
interacting with multiple reservoirs.  The formalism is used to investigate atomic
transport across a variety of lattice configurations.  We demonstrate how the
behavior of an electronic diode, a field-effect transistor, and a bipolar
junction transistor can be realized with neutral, ultracold atoms trapped in
optical lattices.  An analysis of the current fluctuations is provided for the
case of the atomtronic diode.  Finally, we show that it is possible to
demonstrate AND logic gate behavior in an optical lattice.  \end{abstract}

\maketitle

\section{INTRODUCTION}

The emerging field of atomtronics \cite{Zoller,Zozula,myprl} aims to construct
analogies of electronic components, systems and devices using ultracold atoms.
In atomtronics, ultracold atoms move in an optical or magnetic potential in
direct analogy with electrons moving in a semiconductor crystal.  The
motivation to construct and study atomtronic analogs of electronic systems
comes from several directions.

First, the experimental atomtronic realizations promise to be extremely clean.
Imperfections such as lattice defects or phonons can be completely eliminated.
This allows one to study an idealized system from which all inessential
complications have been stripped.  Consequently, one may obtain an improved
understanding of the essential requirements that make certain electronic
devices work. It is possible that a deeper understanding may feed back to the
design of conventional electronic systems and could lead to future
improvements. This lies parallel to the recent interest in
single electron transistors in mesoscopic systems \cite{AZG} and molecules
\cite{ArizonaGang}, where many themes common with atomtronics emerge.  A
consequence of the near-ideal experimental conditions for optical lattice
systems is that theoretical descriptions for atomtronic systems can be
developed from first principles. This allows theorists to develop detailed
models that can reliably predict the properties of devices.

Second, atomtronics systems are richer than their electronic counterparts
because atoms possess more internal degrees of freedom than electrons. Atoms
can be either bosons or fermions, and the interactions between atoms can be
widely varied from short to long range and from strong to weak.  This can lead
to behavior that is qualitatively different to that of electronics
\cite{Theis:opticalFR,Inouye:MoleculeFR,Inouye:heteromolecules,Stan:heteromolecules}.
Consequently, one can study repulsive, attractive, or even non-interacting
atoms in the same experimental setup.  Additionally, current experimental
techniques allow the detection of atoms with fast, state-resolved, near unit
quantum efficiency \cite{GreinNat}.  Thus it is possible, in principle, to
follow the dynamics of an atomtronic system in real time.

Third, neutral atoms in optical lattices can be well-isolated from the
environment, reducing decoherence.  They combine a powerful means of state
readout and preparation with methods for entangling
atoms\cite{Mandel:entanglementlattices}. Such systems have all the necessary
ingredients to be the building blocks of quantum signal processors. The close
analogies with electronic devices can serve as a guide in the search for new
quantum information architectures, including novel types of quantum logic
gates that are closely tied with the conventional architecture in electronic
computers.

Fourth and finally, recent experiments studying transport properties of
ultracold atoms in optical lattices
\cite{EsslingerMoritz,RaizenLi06,PhillipsRolston02} can be discussed in the
context of the atomtronics framework.  In particular, one can model the
short-time transport properties of an optical lattice with the open quantum
system formalism discussed here. 

In this article we present a derivation of the master equation used to treat
these specific open quantum systems.  Afterward, we provide a detailed
analysis of atomtronic analogies of the most elementary electronic components.
These include conducting wires, diodes, and transistors.  This work builds on a
previous paper \cite{myprl}, providing a comprehensive explanation of the
underlying analytical and numerical methods, and additional analysis of the
components.  Finally, we propose how AND logic gate behavior can be recovered
in this open quantum system setting.

\begin{figure}
\includegraphics[width=.7\columnwidth]{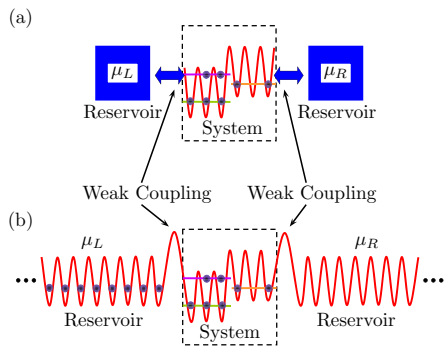}
\caption{(Color online.) The illustration of an optical lattice, energetically
tailored to have a transport response of an electronic diode, connected to two
reservoirs whose respective chemical potentials induce an atomic transport from left
to right.}
\label{ArbLattice}
\end{figure}

This article is organized as follows.  In Sec. \ref{MasterEqn} we discuss the
general master equation formalism, introducing the specific systems to be
investigated, the defining properties of the reservoir, and the
appropriate approximations necessary to complete the derivation for the models.
In Secs. \ref{AtAps} and \ref{Logic} we apply the derived model to a variety of
one-dimensional optical lattice systems in different open quantum system
settings.  The result is a collection of atomtronic devices that emulate
electronic components.

\section{MODEL\label{MasterEqn}}

\subsection{System}

Unlike the typical behavior of their electronic counterparts, atomtronic
devices operate at the few-particle level. The necessary repulsive correlations
between particles are generally caused by either strong interactions for
bosons, or quantum statistical effects due to the Pauli exclusion principle for
fermions. In this article we focus on bosonic current carriers, partly to draw
contrast to the electronic case.

The optical lattice provides a clean, controllable environment for atomtronics
components.  Strong atom-atom interactions can be precisely tuned if the atoms
are confined in a tight optical lattice. Holding the atoms in a lattice has two
primary advantages. First, the strength of the interactions is enhanced due to
the confinement within the lattice wells and second, by cooling the atoms deep
into the lowest Bloch band regime, the center of mass motion of the atoms can
be reduced to hopping between neighboring lattice sites. This results in a
simple theoretical description as well as a clean experimental realization.

The dynamics of these ultracold atoms in this regime are accurately described
by the Bose-Hubbard model \cite{JakchZoller,Greiner}.  In the lowest Bloch
band, the Bose-Hubbard Hamiltonian is
\begin{equation}
\begin{split}
  \hat{H}_{\rm{BH}} =\sum_{i=1}^{N}\bigg(\epsilon_{i}\hat{N}_{i}+\frac{1}{2}U_{i}\hat{N}_{i}\left( \hat{N}_{i}-1\right)\bigg)  \\
  +\sum_{\left\langle i,j\right\rangle }\bigg(J_{ij} \a_{i}^{^{\dag
      }}\a_{j} + \rm{H.c.}\bigg) ,
\label{eqnBH}
\end{split}
\end{equation} 
where $i$ and $j$ are lattice site indices, $N$ is the total number of lattice
sites, $\epsilon _{i}$ is the site energy, $U_{i}$ is the on-site interaction
energy,  $J_{ij}$ is the hopping energy, the sum $\langle i,j\rangle$ is taken
between adjacent lattice sites, and $\hat{N}_{i}=\a_{i}^{\dag }\a_{i}$.  Here,
$\a_{i}^{\dag }$  and $\a_{i}$ are bosonic creation and annihilation operators
respectively for an atom in the lowest Wannier orbital centered at site $i$.
Due to the precise tunability of the experimental parameters, $\epsilon_i$,
$U_{i}$, and $J_{ij}$ in the laboratory today, we use the Bose-Hubbard
Hamiltonian to model our custom lattice atomtronic systems.

\subsection{Reservoir} The reservoirs serve two purposes in atomtronics: they
are `sources' and `sinks' of particles. The reservoirs themselves could be
experimentally realized in a variety of different ways.  For instance, a
reservoir `source' of atoms could be a 1-dimensional optical lattice (Fig.
\ref{ArbLattice}), or a strongly-interacting harmonically trapped ultracold
gas. A reservoir `sink' of atoms could be a coupling of the system to vacuum
modes of atoms in an untrapped state or the densely spaced modes of a nearly
empty potential well.  For modeling purposes, suitable reservoirs must meet
several requirements. The system-reservoir correlations must decorrelate from
the system faster than the time scale over which the state of the system
changes appreciably. This allows one to make the Markov approximation. We will
also assume that system and reservoirs are weakly coupled so that we can make
the second-order Born approximation. In order for the reservoirs to serve as
incoherent sources of atoms, their density of states must vary slowly over the
spectrum of the system.

In the model used in this work, it is assumed that the reservoirs are
strongly-interacting bosons and at low temperatures, the states in the
reservoir are filled to the chemical potential, and all states above the
chemical potential are empty. This is analogous to the situation in a
semiconductor crystal, where to a good approximation the electrons occupy all
states up to the Fermi energy.  Labeling the single particle excitations of the
reservoir by a quantum number $k$,  we can write the Hamiltonian of the
reservoir as \begin{equation} \H_R=\sum_k\varepsilon_k\R_k^\dag\R_k,
\end{equation} where  $\varepsilon_k$ and $\R_k^\dag$ are the energy and
creation operator for the $k$'th reservoir mode, respectively.

Such a reservoir can be constructed out of bosons trapped in a deep optical lattice
potential. If the interactions between the atoms are very strong the system
enters the Mott-insulator regime and by adjusting the chemical potential
one can achieve a situation were each lattice site is occupied by precisely one
atom. The atoms in this system could then be coupled to the atomtronic device
to furnish an incoherent source of
particles, as depicted in Fig.~\ref{ArbLattice}. By arranging the distribution of the energies $\varepsilon_k$ in the
reservoir it is possible to achieve a situation reminiscent of the Fermi-sea
reservoir.  We may then represent the reservoir by a collection
of uncoupled harmonic oscillators.

\subsection{Elimination of the reservoir: master equation formulation}

The quantum master equation approach is often used in quantum optics for describing open
quantum systems \cite{Cohen,Meistre,Walls,Zubarev}.  In essence, it allows one to calculate
observables associated with the evolution of a closed system without having to
account for the free evolution of the reservoir. In this section we provide a
detailed derivation of the master equation formulation for the system and
reservoir described above. The derivation involves deriving an equation of
motion for the reduced density operator for the system.  Here we construct this
equation of motion in a Liouville representation, since (as we discuss below)
this form makes clear a way to go beyond the Born approximation, a necessity in
this zero-temperature, unconventional setting.

The coupling of the system to the reservoir is by means of the exchange of
particles between the system and the reservoir. The Hamiltonian for this interaction is
\begin{equation}
\H_V=\sum_{k,q} g_{kq}\R_k^\dag\a_q +\rm{H.c.}, 
\end{equation}
where $\hat a_q$ is the annihilation operator for a particle in a system
lattice site $q$ and  $g_{kq}$ is the coupling matrix element between reservoir
mode $k$ and site $q$.

The Hamiltonian for the system-reservoir interaction is given by 
\begin{equation}
\H=\H_S+\H_R+\H_V.
\label{FullH}
\end{equation}
where $\H_S=H_{BH}$ for our model.  For each part of the Hamiltonian we introduce an operation $\L_x$,
defined by its action on an arbitrary operator $\oTheta$ by
\begin{equation} 
i\L_x\oTheta=\frac{1}{i\hbar}[\oTheta,\H_x], \quad \rm{where} \quad\quad \it{x}\in\{\rm{S,\;R,\;V\}}.
\end{equation}
We denote the density matrix of the system and reservoir with
$\hat\rho$.  From the full density matrix, the reduced density matrices $\hat
\sigma_S$ and $\hat \sigma_R$ of system and reservoir are defined by tracing
over the reservoir and system Hilbert spaces, respectively,
\begin{equation}
\hat \sigma_{S,R} = \Tr_{R,S}[\hat \rho].
\end{equation}
We define the projection operator $\bbP$ by
\begin{equation}
\bbP \oTheta \equiv \osigma_R \otimes\Tr_R[\oTheta]
\end{equation}
and its compliment by $\bbQ =\eye-\bbP$. Under these
definitions,  $\bbP$ and $\bbQ$ satisfy the usual projection operator
relationships $\bbP^2\oTheta=\bbP\oTheta$,
$\bbQ^2\oTheta=\bbQ\oTheta$, and $\bbP\bbQ\oTheta=0$. Using the projection
operators, the reduced density matrix for the system can be written as
\begin{equation}
\osigma_S = \Tr_R[\bbP\orho].
\end{equation}

To find the equation of motion for $\osigma_S$, we start from the evolution of the full
density matrix $\orho$: \begin{equation}
\frac{d\orho}{dt}=-i\L\orho,
\label{Heis}
\end{equation}
where $\L=\L_S+\L_R+\L_V$.  Noting that $\bbP+\bbQ=\eye$,
this equation can be written as 
\begin{equation}
\frac{d\left(\bbP+\bbQ\right)\orho}{dt}=-i\left(\bbP+\bbQ\right)\L\left(\bbP+\bbQ\right)\orho.
\end{equation}
Acting with $\bbP$ and $\bbQ$ separates this equation into the coupled equations
\begin{eqnarray}
\frac{d\bbP\orho}{dt} & = & -i\left[\bbP\L\bbP\orho+\bbP\L\bbQ(\bbQ\orho)\right] \label{projrho1},\\
\frac{d\bbQ\orho}{dt} & = & -i\left[\bbQ\L\bbP\orho+\bbQ\L\bbQ(\bbQ\orho)\right] \label{projrho2}.
\end{eqnarray}
The solution of equation (\ref{projrho2}) is
\begin{equation}
\bbQ\orho=-i\int_0^t  {\rm e}^{-i\bbQ\L\bbQ \tau}\bbQ\L\bbP\orho(t-\tau)d\tau.
\end{equation}
where we have assumed that the system and reservoir initially uncorrelated, {\em i.e.} $\bbQ\orho(0)=0$.  Using this result to eliminate $\bbQ \orho$ in
Eq.~(\ref{projrho1}) yields \begin{equation}
\frac{d\bbP\orho}{dt} = -i\bbP\L\bbP\orho-\bbP \L\bbQ\int_0^t  {\rm e}^{-i\bbQ \L\bbQ \tau}\bbQ  \L\bbP\orho(t-\tau)d\tau.
\end{equation}
Tracing over the reservoir Hilbert space leads to the equation of motion for
the reduced density operator of the system, \begin{equation}
\label{FullMeq}
\frac{d\osigma_S(t)}{dt} = -i\L_S\osigma_S(t) +\int_0^t{\mathscr R}(\tau)\osigma_S(t-\tau)d\tau,
\end{equation}
where we have introduced the memory kernel
\begin{equation}
\label{MemKern}
{\mathscr R}(\tau)=-\Tr_R\left[ \L_V
{\rm e}^{-i(\L_S+\L_R+\bbQ\L_V\bbQ )\tau}\L_V\osigma_R\right].
\end{equation}
The first term on the right hand side of Eq.~(\ref{FullMeq}) is the
free system evolution while the second term describes the
irreversible contribution due to the system-reservoir interaction. 

Equation~(\ref{FullMeq}) is the exact master equation under the condition that
the system and reservoirs are initially uncorrelated. It achieves our goal, in
principle: all the dynamics due to the coupling to the reservoirs is
encapsulated in $\mathscr R(\tau)$. Once the memory kernel is known we can
calculate the evolution of the reduced density matrix of the system without
having to take into account the reservoirs explicitly. However, it turns out to
be impossible to solve for the memory kernel exactly in even the simplest of
circumstances, so approximations must be made to continue from this point.

Calculation of the memory kernel and its action on the system density matrix is
straight forward under the Markov and Born approximations. The Markov
approximation assumes that the correlation time $\tau_c$ between system and
reservoir is much shorter than the time scales over which the reduced density
matrix of the system changes appreciably, {\em i.e.}
\begin{equation}
\tau_c \frac{d \osigma_S}{dt}\ll \osigma_S.
\end{equation}
The Markov approximation consists of treating the system density matrix as a
constant over time intervals of order $\tau_c$, and accordingly we can pull it
out of the integral in Eq.~(\ref{FullMeq}). The short correlation time also
allows us to extend the limit of integration in Eq.~(\ref{FullMeq}) to
infinity.

The Born approximation takes the memory kernel to second order in $\L_V$ and is
thus a weak coupling approximation. In the conventional Born approximation,
$\L_V$ is eliminated from the exponential term in the memory kernel.  We cannot
simply employ such an approximation here however since we are assuming that the
strongly-interacting boson gas obeys a zero temperature Fermi-Dirac
distribution characterized by $\mu_F$.  The hard edge at the Fermi energy
causes logarithmic divergences in the second-order energy shifts of system
levels as $\mu_F$ approaches system resonances.  This divergence is due to the
fact that, under the second-order approximation, as $\mu_F$ approaches a system
resonance, the number of reservoir modes that the system is coupled to goes to
one---not a continuum of modes.  This would induce Rabi flopping of the atom
between the system and the reservoir and not the irreversible reservoir action
intended.  In reality, the interaction of the system with the reservoir taken
to all orders mixes the modes and leads to decay, in this situation.  Exact
simulations on small systems show that we can recover the proper dynamics at
the hard edge by including the influence of higher order terms of the memory
kernel expansion in $\L_V$.  We find that the fourth order term in the
expansion provides a good estimate for the higher order corrections of the full
memory kernel, yielding \begin{equation} \label{mkern} {\mathscr
R}(\tau)\approx  -\Tr_R\left[ \L_V \L_V(-\tau)\osigma_R {\rm
e}^{-[i(\L_S+\L_R)+\eta] \tau}\right] , \end{equation} where $\L_V(\tau)={\rm
exp}[-i(\L_S+\L_R)\tau]\L_V$.  The evaluation of the fourth order term shows
that the decay $\eta$ has a small dependence on the value $\mu_F$ about the
system eigenenergy difference.  Here, we take $\eta$ to be the mean value of
the decay rate, which is where $\mu_F$ equals the system eigenenergy
difference, that is $\eta={\mathscr D}J^2\pi/(2\hbar^2)$, where ${\mathscr D}$
is the density of reservoir modes, assumed to be constant in the region of
interest.  Introducing the rate at which particles enter or leave the system
$\Gamma_0\equiv{\mathscr D}J^2/\hbar^2$, we have  $\eta=\pi\Gamma_0/2$.  Taking
this value for $\eta$ is reasonable because $\eta$ is only important in the
calculation for only small deviations.  The modified memory kernel in
Eq.~(\ref{mkern}) captures the correct long-time behavior of the exact memory
kernel.

With these approximations the master equation becomes
\begin{equation}
\label{BMMeq}
\frac{d\osigma_S}{dt} \approx -i\L_S\osigma_S - \int_0^\infty {\rm e}^{-\eta
\tau}\Tr_R\left[ \L_V \L_V(-\tau)\osigma_R\right] d\tau\osigma_S(t).
\end{equation}
Inserting the Liouvillians into Eq.~(\ref{BMMeq}) yields
\begin{equation}
\label{BMMeq2}
\begin{split}
\frac{d\osigma_S}{dt} \approx & -i\L_S\osigma_S - \frac{1}{\hbar^2}\sum_k |g_{kq}|^2 
\int_0^\infty d\tau {\rm e}^{-\eta\tau} \times \Bigg(\\
& \Tr_R\bigg\{
\Big[\hat a_q^\dag \R_k,[\hat a_q(-\tau) \R^\dag_k(-\tau),\osigma_S\osigma_R]\Big] \bigg\}\\ 
&+\Tr_R\bigg\{\Big[\hat a_q \R^\dag_k,[\hat a_q^\dag(-\tau) \R_k(-\tau),\osigma_S\osigma_R]\Big]
\bigg\} \Bigg),
\end{split}
\end{equation}
where we have used that
\begin{equation}
\begin{split}
{\rm Tr}_R[\osigma_R \hat{R}^\dagger_k\hat{R}_{k^\prime}] = \delta_{k,k^\prime}\langle \hat R_k^\dagger \hat R_{k}\rangle,\\
{\rm Tr}_R[\osigma_R \hat R_k \hat R_{k^\prime}^\dagger] = \delta_{k,k^\prime}\langle \hat R_k\hat R_{k}^\dagger\rangle,\\
\end{split}
\end{equation}
due to the fact that the reservoir is in thermal equilibrium. 

We project Eq.~(\ref{BMMeq2}) onto the energy eigenbasis of the system and
trace out the reservoir degrees of freedom. This allows us to evaluate the
$\tau$ integral and to find a more explicit form of the master equation. Given
two arbitrary system energy eigenstates $\ket{a}$ and $\ket{b}$ and adopting
the notation
$\bra{a}\hat{\Theta}\ket{b}=\langle\hat{\Theta}\rangle_{ab}$, we have 
\begin{equation}
\begin{split}
\langle a |\L_S\osigma_S |b\rangle=\omega_{ab}\langle\osigma_S\rangle_{ab} ,\\ \langle
a|\hat{a}^\dag(-\tau)|b \rangle={\rm
exp}(-i\omega_{ab}\tau)\langle\hat{a}^\dag\rangle_{ab} 
\end{split}
\end{equation}
with
$\hbar\omega_{ab}=(E_a-E_b)$ being the difference between the eigenenergies of
$\ket{a}$ and $\ket{b}$. Similarly,
\begin{equation}
\langle
i|\hat{R}^\dagger(-\tau)|j \rangle={\rm
exp}(-i\omega_{ij}\tau)\langle\hat{R}^\dagger\rangle_{ij}
\end{equation}
for the reservoir. Performing the
integral over $\tau$ produces the following closed form of the master equation:
\begin{widetext}
\begin{eqnarray}
\label{BMMeq3}
\frac{d\langle \osigma_S\rangle_{ab}}{dt} &=&-i\omega_{ab}\langle \osigma_S\rangle_{ab} -\frac{1}{2}\sum_{c,d}\bigg\{
\langle\a_q^\dag\rangle_{ac}\Big[(\Gamma^{({\rm Out})}_-)_{cd}\langle\a_q\rangle_{cd}\Big]\langle \osigma_S\rangle_{db}
-\Big[(\Gamma^{({\rm Out})}_-)_{ac}\langle\a_q\rangle_{ac}\Big]\langle \osigma_S\rangle_{cd}\langle\a_q^\dag\rangle_{db}\nonumber\\
&&+\langle \osigma_S\rangle_{ac}\Big[(\Gamma^{({\rm In})}_-)_{cd}\langle\a_q\rangle_{cd}\Big]\langle\a_q^\dag\rangle_{db}
-\langle\a_q^\dag\rangle_{ac}\langle \osigma_S\rangle_{cd}\Big[(\Gamma^{({\rm In})}_-)_{db}\langle\a_q\rangle_{db}\Big]
+\langle\a_q\rangle_{ac}\Big[(\Gamma^{({\rm In})}_+)_{cd}\langle\a_q^\dag\rangle_{cd}\Big]\langle \osigma_S\rangle_{db} \nonumber\\
&&-\Big[(\Gamma^{({\rm In})}_+)_{ac}\langle\a_q^\dag\rangle_{ac}\Big]\langle \osigma_S\rangle_{cd}\langle\a_q\rangle_{db}
+\langle \osigma_S\rangle_{ac}\Big[(\Gamma^{({\rm Out})}_+)_{cd}\langle\a_q^\dag\rangle_{cd}\Big]\langle\a_q\rangle_{db}
-\langle\a_q\rangle_{ac}\langle \osigma_S\rangle_{cd}\Big[(\Gamma^{({\rm Out})}_+)_{db}\langle\a_q^\dag\rangle_{db}\Big]\bigg\} ,
\end{eqnarray}
\end{widetext}
where
\begin{eqnarray}
(\Gamma_\pm^{({\rm In})})_{ab}&=&\frac{2}{\hbar^2}\sum_k \frac{|g_{kq}|^2}{\eta+i(\pm \omega_k -\omega_{ab})}\langle \hat R_k^\dagger\hat R_k\rangle,\\
(\Gamma_\pm^{({\rm Out})})_{ab}&=&\frac{2}{\hbar^2}\sum_k \frac{|g_{kq}|^2}{\eta+i(\pm \omega_k -\omega_{ab})}\langle \hat R_k\hat R_k^\dagger\rangle.
\end{eqnarray}
Note that 
\[
\lim_{\eta\to 0}\frac{1}{\eta+i(\pm \omega_k -\omega_{ab})}= 
-i {\mathcal P}\frac{1}{\pm \omega_k -\omega_{ab}}+\pi\delta(\pm \omega_k -\omega_{ab}),
\]
where $\mathcal P$ indicates that integrals are to be interpreted in the Cauchy
principal value sense. The real parts of the $\Gamma_\pm$ matrices then give decay
rates that agree with the Fermi golden rule result while the imaginary parts
give rise to level shifts in analogy with the Lamb shift in the hydrogen
spectrum.

\subsection{Reservoir model}
The detailed physics of the reservoirs influence the evolution of the system
through the coupling matrix elements $g_{kq}$, the occupation probabilities of
the reservoir modes $\langle \hat R_k^\dagger \hat R_k\rangle$, and the density
of states of the reservoir. The reservoir model used below assumes fermionized
strongly-interacting bosons.  Thus, $\langle \hat R_k^\dagger \hat
R_k\rangle=1$ for all modes below the chemical potential of the reservoir and
$\langle \hat R_k^\dagger \hat R_k\rangle=0$ for all modes above.  The coupling
of the reservoir modes to the system states is a slowly varying function of the
mode energy. We model it by a constant coupling up to some high energy
cut-off $\omega_c$ in order to quench the ultra violet divergence that would
otherwise arise, taking the form,
\begin{equation}
|g_{kq}|^2= |g_q|^2\theta (\omega_c-\omega_k),
\end{equation} 
with $\theta(\nu)$ the Heaviside step-function. The high energy cut-off
$\omega_c$ is much larger than any relevant frequency of the system, and
it does not affect the system's dynamics.

\section{ATOMTRONICS APPLICATIONS \label{AtAps}}

In order to analyze the response of specific optical lattice configurations connected to
reservoirs with different chemical potentials, we consider the steady-state
solution of Eq.~(\ref{BMMeq3}) and then proceed to solve for the matrix
elements $\langle \osigma_S\rangle_{ab}$. Once $\langle \osigma_S\rangle_{ab}$
is known, expectation values of atomic currents $\langle\hat{J}_q\rangle$ into
(and out of) a system site $q$ can be calculated from reservoir's influence of the system population
rates.  For the population of each state,
$d\langle \osigma_S\rangle_{aa}/dt$ we sum up net rates out of the system state
$\ket{a}$ and then subtract the net rates into the state.  We then sum over all
of the system states $\ket{a}$ to obtain the net current transport on system
site $q$:
\begin{widetext}
\begin{eqnarray}
\langle\hat{J}_q\rangle&=&\sum_a\sum_{c,d}\bigg\{
\langle\a_q^\dag\rangle_{ac}\Big[(\Gamma^{(Out)}_-)_{cd}\langle\a_q\rangle_{cd}\Big]\langle \osigma_S\rangle_{da}
+\langle \osigma_S\rangle_{ac}\Big[(\Gamma^{(Out)}_+)_{cd}\langle\a_q^\dag\rangle_{cd}\Big]\langle\a_q\rangle_{da}\nonumber\\
&&\qquad\qquad-\langle\a_q^\dag\rangle_{ac}\langle \osigma_S\rangle_{cd}\Big[(\Gamma^{(In)}_-)_{da}\langle\a_q\rangle_{da}\Big]
-\Big[(\Gamma^{(In)}_+)_{ac}\langle\a_q^\dag\rangle_{ac}\Big]\langle \osigma_S\rangle_{cd}\langle\a_q\rangle_{da}\bigg\}\\
&=&\sum_a\sum_{d}\langle\hat{J}\rangle_{ad}\langle \osigma_S\rangle_{da}\\
&=&{\rm Tr}_S\Big[\hat{J}_q\osigma_S\Big].
\end{eqnarray}
\end{widetext}
where
\begin{widetext}
\begin{equation}
\label{jumpcurrent}
\langle\hat{J_q}\rangle_{ad}=\sum_c\bigg\{
\langle\a_q^\dag\rangle_{ac}\Big[(\Gamma^{(Out)}_-)_{cd}\langle\a_q\rangle_{cd}\Big]
+\Big[(\Gamma^{(Out)}_+)_{ac}\langle\a_q^\dag\rangle_{ac}\Big]\langle\a_q\rangle_{cd}
-\Big[(\Gamma^{(In)}_-)_{ac}\langle\a_q\rangle_{ac}\Big]\langle\a_q^\dag\rangle_{cd}
-\langle\a_q\rangle_{ac}\Big[(\Gamma^{(In)}_+)_{cd}\langle\a_q^\dag\rangle_{cd}\Big]\bigg\}.
\end{equation}
\end{widetext}
as the current operator for site $q$ projected onto the system eigenbasis.
Using this convention, the sign of $Tr_S[\hat{J}_q\osigma_S]$ reveals whether
the current flows into ($-$) or out of ($+$) the system.

\subsection{Atomtronics analogy of a simple circuit}
\begin{figure}
\includegraphics[width=.8\columnwidth]{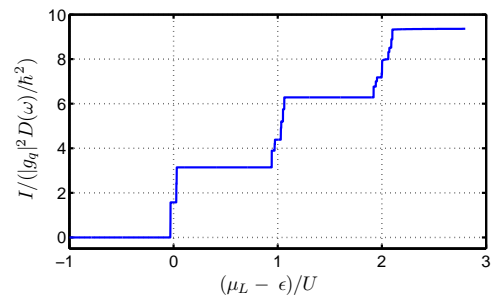}
\caption{(Color online.) Current of the atomtronic wire, in the
case of weak coupling to the reservoirs, as a function of the left chemical
potential.  The current monotonically increases in quantized jumps as the
on-site interaction energy to put on an additional particle is overcome. The
parameters used to model the atomtronic wire in this simulation are
$\epsilon/U=3$, $J/U=3\times 10^{-2}$, and $\hbar\Gamma_0/U=10^{-6}$. }
\label{WireData}
\end{figure}

Here we analyze the atomtronics counterpart to a simple circuit of a
battery connected to a resistive wire.
\begin{figure}
\includegraphics[width=.8\columnwidth]{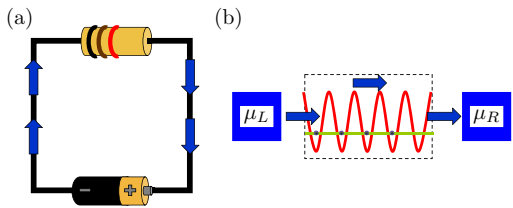}
\caption{(Color online.) Illustration of the analogy between (a) the electronic
circuit of a wire (with some inherent resistance) connected to a battery and (b)
its atomtronic counterpart.} \label{WireAnalogy}
\end{figure} 
As seen in Fig. \ref{WireAnalogy}, the analogy of a wire is an
energetically-flat optical lattice, with uniform tunneling rates and
interaction energies ( $\epsilon_j\equiv\epsilon$,  $U_j\equiv U$, and
$J_{ij}\equiv J$ for all neighboring sites).   For this system, that is very
weakly-coupled to the reservoirs, we calculate the atomic current as a function
of chemical potential difference.  This numerical experiment is carried out by
initially setting both left and right chemical potentials ($\mu_L$ and $\mu_R$)
to zero.  We raise $\mu_L$ so that an atomic transport is induced across the
system from left to right, we compute the current $\langle J_R\rangle$ out of
the right side of the system shown in Fig. \ref{WireData}.  The current
increases with the chemical potential difference, but in quantized jumps that
correspond directly to the left chemical potential overcoming the on-site
interaction energy needed to introduce a greater number of atoms onto the left
site. A closer examination of the numerical simulation implemented in Fig.
\ref{WireData} reveals two subtle features.  Moving from left to right across
graph in Fig. \ref{WireData}, the first current jump, occurring at
$(\mu_L-\epsilon)/U=0$, the current increases in two steps where one might
expect to observe a single jump, since the condition to put a particle in the
lattice is $\mu_L\geq\epsilon$.  This is a result of the fact that the
degeneracy of the Fock states in the one-particle manifold are split by $2J$ in
the system state eigenbasis.  In addition the jump in current is broadened
slightly by the nonzero $\eta$ and the system-reservoir coupling.  This
broadening, which makes the jump in current a smooth transition is more
apparent in the atomtronic devices presented below where the system
reservoir-coupling is taken to be orders of magnitude larger.  Although the
exact details for the second and third jumps are more complicated, the
reasoning is the same: the eigenenergies are split by approximately $2J$, and
the overall jump is smoothed out by $\eta$.  These are general properties of
all of the numerical experiments described in this work.

\subsection{The atomtronic diode} 
A diode is a device with an approximately unidirectional current
characteristic.  A voltage bias across the diode yields a current in one
direction but not in the opposite direction if the voltage bias is reversed.
Such behavior can be realized in an optical lattice by creating an energy shift
in half of the lattice with respect to the other.  We find that the diode
characteristic persists as the number of lattice sites is increased.  For
simplicity, here we present the diode in a two-site lattice system.  For the
simulations in the rest of this paper, we assume $\epsilon/U=3$, $J/U=3\times
10^{-2}$, $\Gamma_0\hbar/U=10^{-2}$.

In the Fock basis for a two site system, there exist three states in the
two-particle manifold: $|2 0\rangle$,  $|1 1\rangle$, and $|0 2\rangle$, where
$|n m\rangle$ refers to $n$ particles on the left site and $m$ particles
on the right.  The external energies of the two sites ($\epsilon_1$ and
$\epsilon_2$) can be chosen so that the eigenstates $|2 0\rangle$ and  $|1
1\rangle$ approximately degenerate, leaving both states far detuned from $|0 2
\rangle$.  This configuration of the site energies is given by
$\epsilon_2=\epsilon_1 + U$.  We refer to this as the ``resonance condition''. 

\begin{figure}
\includegraphics[width=.8\columnwidth]{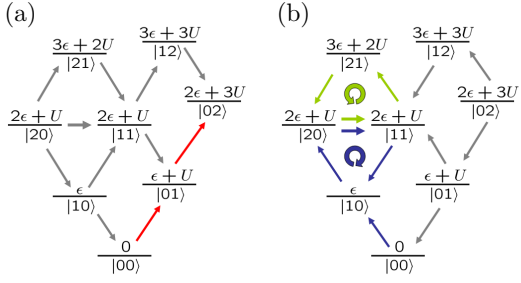}
 \caption{(Color online.) (a) Energy schematic of the reverse bias dynamics of the two site
atomtronic diode.  The red arrows represent the system transitions from an
initially empty system.  The gray arrows represent all other possible
transitions.  Regardless of which state the system starts in, it evolves almost
entirely to the $|02\rangle$ state. (b) Energy schematic of the forward-bias
dynamics of the two-site atomtronic diode. Blue and green arrows illustrate
current-bearing cycles, while gray arrows represent all other possible
transitions.} \label{fig:diode} 
\end{figure}

Figure \ref{fig:diode} illustrates how the resonance condition generates
reverse-bias and forward-bias behavior in a two-site optical lattice.  As seen
in Fig.  \ref{fig:diode}(a), if one holds the left reservoir chemical potential at
$\mu_L=0$ and raises the right reservoir chemical potential $\mu_R$, the system
will undergo a transition from $|0 0\rangle$ to $|0 1\rangle$.  The states $|0
1\rangle$ and $|1 0\rangle$ are separated in energy by $U$.  As a result, most
of the population remains in the $|0 1\rangle$ state.  Increasing $\mu_R$ above
the point where the transition from $|0 1\rangle$ to $|0 2\rangle$ is allowed,
the system remains almost completely settled in the $|0 2\rangle$ state.

As seen in the Fig. \ref{fig:diode}(b), if one holds $\mu_R=0$ and raises
$\mu_L$, the system first undergoes a transition from $|0 0\rangle$ to $|1
0\rangle$.  However, increasing $\mu_L$ so the system evolves to $|2 0\rangle$
leads to a very different situation than in the above case: since $|2 0\rangle$
is resonant with $|1 1\rangle$, both states are simultaneously populated.
Since $\mu_R=0$ takes all particles out of the site on the right, the system
can make a transition from $|1 1\rangle$ back to $|1 0\rangle$. The combined
effect of setting $\mu_L$ and $\mu_R$ to these values is to force the system to
undergo a closed cycle of transitions between $|1 0\rangle$, $|2 0\rangle$ and
$|1 1\rangle$.  The result  is a net atomic transport (or current flow) across
the system.  A second contributor to the net current through the system is the
fact that $\mu_L$ allows transitions from $|1 1\rangle$ to $|2 1\rangle$.
Thus, an additional current-generating cycle exists: $|1 1\rangle$ to $|2
1\rangle$ to $|2 0\rangle$ and back to $|1 1\rangle$. Both cycles contribute
positively to a net current flow across the system.

\begin{figure}
\includegraphics[width=.7\columnwidth]{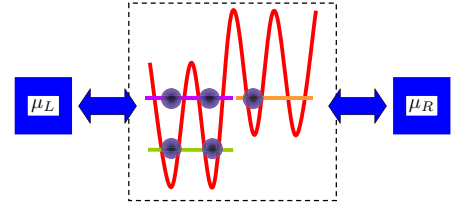}
\caption{(Color online.) Schematic of the four-site optical lattice atomtronic
diode.} \label{fig:diodeSchem}
\end{figure}

For systems consisting of $N$ lattice sites, the diode configuration consists of 
two connected, energetically-flat lattices whose energy separation is
$\Delta\epsilon=U$. The dynamics do not change for larger systems since the degeneracy of the flat lattice allows for effective transport across the lattice,
allowing a particle to enter the left site of the  half of the lattice is
energetically degenerate with finding a particle at the junction.  Thus, there
exist current cycles initially generated from  the $|222...2\rangle\otimes
|0...000\rangle$ to $|222...1\rangle\otimes |1...000\rangle$ transition.  Going
the other direction, one can go to $|0...000\rangle\otimes |222...2\rangle$,
but conditions are not energetically favorable to allow atomic transport across
the junction. Figure \ref{fig:diodeSchem} is a schematic of a four-site
atomtronic diode in the forward-biased direction.

\begin{figure}
\includegraphics[width=\columnwidth]{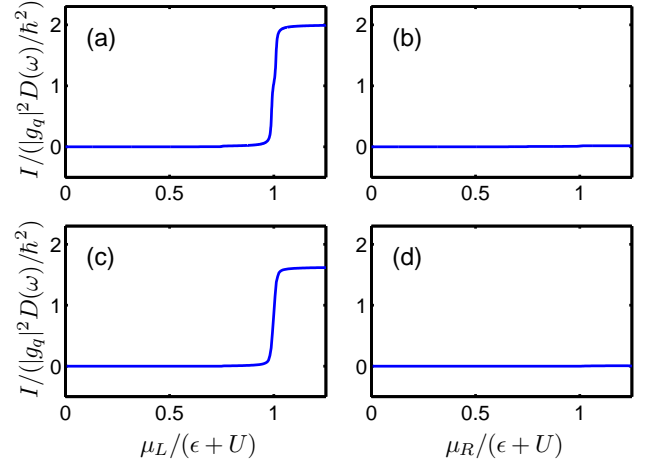}
\caption{(Color online.) Current responses of (a)
forward-bias two site diode, (c) the four site diode as a function of
$\mu_L$ (with $\mu_R=0$); current response of (b) the
reverse-bias two site diode and (d) four site diode as a function of
$\mu_R$ (with $\mu_L=0$).  The chemical potentials of all figures are
normalized to the resonance condition, and the currents normalized to $\Gamma_0$.} \label{diodedata}
\end{figure}

Supporting the behavior that the dynamics of the diode are
qualitatively-independent of the overall size of the lattice, Figs.
\ref{diodedata}(a-b) and \ref{diodedata}(c-d) are numerical simulations of the
current responses of the two site, and four site diodes, respectively.  The
general features of both diodes are qualitatively identical.

Given the current characteristics of the diode, one might ask how much does the
signal fluctuate. Recalling the current operator from Eq.~(\ref{jumpcurrent}),
we can calculate the autocorrelation function for the current,
\begin{equation}
\langle\hat{J_q}(\tau+t)\hat{J_q}(t)\rangle = \Tr_S[\hat{J}_q(\tau+t)\hat{J}_q(t)\osigma_S],
\end{equation}
using the quantum regression theorem.

To simulate an actual measurement, we convolve this correlation function with
an exponentially decaying filter, ${\rm exp}(-t/T)$. The Fourier transform of
this convolution yields a time-averaged spectral density function
$S(\omega,T)$.  Our time-averaged signal-to-noise ratio SNR as a function of
$T$ is then
\begin{equation}
{\rm SNR}(T)=\frac{\langle\hat{J_q}\rangle}{\sqrt{\int_0^\infty S(\omega,T)d\omega}} .
\end{equation}
For long averaging times $T$ we find, for the conditions of Fig. \ref{diodedata},
\begin{equation}
{\rm SNR}(T)\approx \sqrt{8\Gamma_0}\sqrt{T}
\end{equation}
For typical optical lattice experiments, $U\sim 1{\rm kHz}$ is achievable,
which implies that $\sqrt{8\Gamma_0}\sim 10 \sqrt{{\rm Hz}}$. Therefore, a
signal-to-noise ratio of $10$ can be achieved by averaging the atomic current
for about $1$ second.

\subsection{The atomtronic field-effect transistor} 
\begin{figure}
\includegraphics[width=.8\columnwidth]{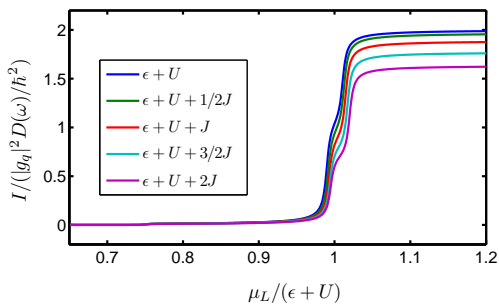}
\caption{(Color online.)
Current response of the atomtronic FET as a function of the chemical potential:
small changes in $\epsilon_2$ lead to appreciable changes in the current
response.  That is, as the system is detuned from its resonance condition,
there is a fall off of the net current value across the device.}
\label{FETdata} \end{figure}

A field-effect transistor (FET) is a device that allows an externally-applied
field to affect the current through the device.  This characteristic allows the
FET to be utilized as an amplifier.  Since the diode is optimized when the
resonance condition is imposed on the optical lattice, small deviations from
the resonance condition lead to large changes in maximum current propagating
across the lattice. This is precisely the behavior of a FET where a current is
controlled by an applied voltage.  In Fig.  \ref{FETdata}, we plot several
current results for the forward-bias configuration as the separation in the
external energy of the second site is raised past the resonance condition by
fractions of $J$, the smallest system parameter in the model.

\subsection{The atomtronic bipolar junction transistor}
A bipolar junction transistor (BJT) is a three terminal device in which the
overall current across the emitter and collector is controlled by a much weaker
current via the base.  Two practical applications of the BJT are signal
amplification and switching (on and off) of the emitter current.

\begin{figure}
\includegraphics[width=\columnwidth]{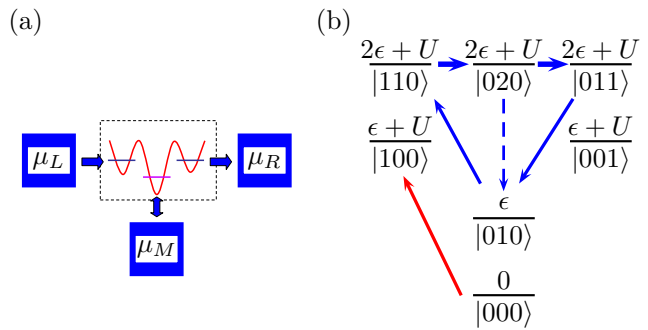}
\caption{(Color online.) (a) An illustration of the BJT lattice-reservoir
system.  (b)  The energy schematic of the three-site optical lattice under the
extended resonance condition.  Here the left reservoir is set to maintain an
occupancy of one atom on the left site and the right is set to remove all
atoms. If the middle reservoir is set to remove all particles, then the system
evolves to the $|100\rangle$ state (red arrow).  If the middle reservoir
is set to maintain an occupancy of one atom on the middle site, then the
degeneracy between states $|110\rangle$,  $|020\rangle$, and $|011\rangle$
allows current to traverse the system.} \label{TransSchem}
\end{figure}

Realization of BJT-like behavior in atomtronic systems requires at least three
sites connected to three different reservoirs.  If the atomtronic diode is
considered an atomtronic p-n junction, one might guess that the atomtronic
n-p-n transistor would entail raising the external energy of the left and right
(collector and emitter) sites higher than the middle (base) site by the on-site
interaction energy.  This configuration is illustrated in Fig.
\ref{TransSchem}(a). 
The reason this configuration yields BJT-like behavior is due to the
approximate degeneracy between the Fock states $|110\rangle$, $|020\rangle$,
and $|011\rangle$.

We implement numerical simulations of this lattice configuration by fixing a
chemical potential difference across the lattice. The left reservoir chemical
potential $\mu_L$ is set to maintain one particle on the left site and the
chemical potential of the right reservoir $\mu_R$ is set to zero.  The middle
chemical potential $\mu_M$ starts at zero and is increased to allow a single
atom to enter the middle site.

When there are no atoms on the middle site, the configuration of the reservoirs
pumps the system into the $|100\rangle$ Fock state (as seen in Fig.
\ref{TransSchem}(b)).  The $|100\rangle$ and the $|001\rangle$ states are
degenerate with each other, but the system must undergo a second-order,
off-resonant transition via the $|010\rangle$ state from $|100\rangle$ to
$|001\rangle$.  Such transitions are suppressed by a factor of  $(J/U)^2$ and
thus become less likely as the energy difference between $|100\rangle$ and
$|010\rangle$ increases.  Thus, when the middle reservoir is set to maintain
zero atoms on the middle site, the net current out of the emitter is minimal.

When the middle reservoir's chemical potential is increased to allow a single
atom into the middle site of the system, the degeneracy between the
$|110\rangle$, $|020\rangle$, and $|011\rangle$ states is accessed, which
allows atoms to travel across the system.  One issue with this configuration is
the following: in order to get to $|011\rangle$, the system has to make a
transition through the $|020\rangle$ state.  Since the middle reservoir is set
to maintain an occupancy of one, but not two, atoms on the middle site, one of
the atoms can be lost to the middle reservoir, leading to a loss of current out
of the emitter.  If the couplings of all three reservoirs to the system are
equal, then the result is the current measured passing through the base turns
out to be even greater than the current measured out of the emitter.   Thus,
the system represents an inefficient transistor realization.  On the other
hand, if the middle reservoir were to be coupled weakly compared to the other
reservoirs, then current predominantly leaves the system via the emitter, which
is the desired behavior.

\begin{figure}
\includegraphics[width=\columnwidth]{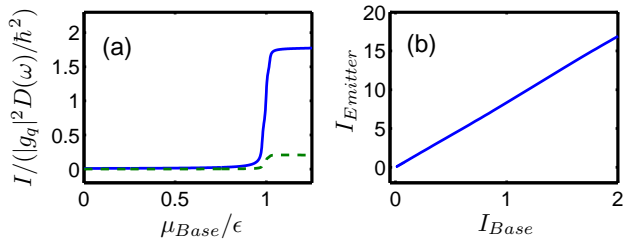}
\caption{(Color online.) Characteristics of the atomtronic BJT. (a) For a fixed
collector-emitter voltage bias, the current response measured through the base
leg (green) and the emitter leg (blue) are plotted as a function of the base
chemical potential.  (b) The current out of the emitter is plotted vs. the
current out of the base.  Here a large linear gain is observed.}
\label{TransData}
\end{figure}

Figure \ref{TransData}(a) is a numerical simulation of the current out of the
emitter and the base as a function of the base chemical potential. The coupling
strength of the base connected to the reservoir is one fifth the collector and
emitter reservoir coupling.  It should be noted that the region where the
proposed atomtronic transistor mimics the electronic BJT is limited to the
transition region, or current jump.  One can increase the length of this region
by increasing the overall system-reservoir coupling strength. Figure
\ref{TransData}(b) shows that the gain of this device is fairly linear.

\section{DISCRETE ATOMTRONICS LOGIC\label{Logic}}
Integrated circuits are designed with a very large number of transistor elements
to perform a desired function.  The demonstrated ability to realize
atomtronic diodes and transistors thus motivates the question as to whether
higher functionality can be realized with these ultracold atomic systems.  Here
we look at the most fundamental of these, the atomtronic AND logic gate.

A traditional logic element is a device with a given number of inputs and
outputs, composed of switches, that generates a series of logical responses.
Such logical behavior can be expressed in a truth table composed of $1$'s and
$0$'s (`ons' and `offs').  Logic elements are the fundamental building blocks of
computing and discrete electronics.  In table \ref{Truth}(a) the truth table
for the AND logic gate is given as an example.  The next level of complexity in
emulating electronic systems is to create logic elements from the atomtronic
components.
  
\begin{figure}
\includegraphics[width=.7\columnwidth]{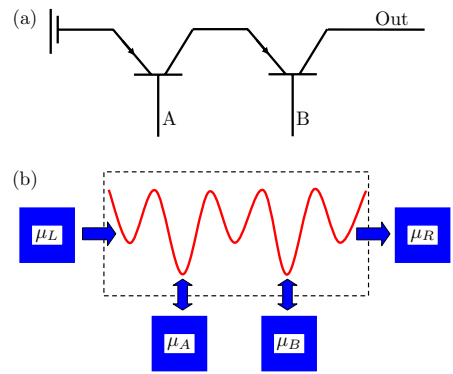}
\caption{(a) Electronic schematic for the AND logic gate, a device constructed
by cascading two transistors in series. (b) Atomtronics AND schematic.  The
atomtronic AND logic gate is constructed exactly like its electronic
counterpart, {\em i.e.} by cascading two transistors in series.  A chemical
potential bias $\mu_L-\mu_R$ across the device attempts to drive a current
across the device.  No substantial current is observed, however, unless both
$\mu_A$ and $\mu_B$ supply atoms onto the base terminals of their respective
transistors.}
\label{and}
\end{figure}

An AND gate is a device with two inputs ($A$ and $B$), and one output ($O$). As
illustrated in table \ref{Truth}(a), the device characteristic of the AND gate
is that $O$ remains off unless both $A$ and $B$ are on.  In electronics, such a
device can be constructed by connecting two transistors in series (as
illustrated in Fig.  \ref{and}(a)).  By analogy, if the atomtronic BJTs are
connected in the same series configuration (as illustrated in Fig.
\ref{and}(b)), the AND gate truth table can be generated.

When constructing practical logic circuits, the values of the $1$s and $0$s are
not strict values, they are defined within a given range. The data in table
\ref{Truth}(b) has been generated in a numerical experiment of the
configuration depicted in Fig. \ref{and}(b).  For this particular experiment,
the maximal current out of the device is at least a factor of $6$ greater than
any other measured current out.  Thus a discernible difference between  `on'
and `off' is observed and the output currents reproduce the AND gate truth
table.  Such a difference can also be enhanced by increasing the on-site
interaction energy $U$ of the lattice.
\begin{table}
\begin{tabular}{|c|c|c|}
\multicolumn{3}{l}{(a)} \\\hline
\multicolumn{3}{|c|}{AND Gate} \\\hline
$A$ & $B$ & $O$ \\\hline
0 & 0 & 0\\\hline
1 & 0 & 0\\\hline
0 & 1 & 0\\\hline
1 & 1 & 1\\\hline
\end{tabular}
\hspace{.5in}
\begin{tabular}{|c|c|c|}
\multicolumn{3}{l}{(b)}\\
\hline
\multicolumn{3}{|c|}{Atomtronic AND Gate Simulation}\\\hline
$A$ & $B$ & $O$ \\
\hline
$0$ & $0$ & $0.00$ \\\hline
$3.2$ & $0$ & $0.01$ \\\hline
$0$ & $3.2$ & $0.16$ \\\hline
$3.2$ & $3.2$ & $1.00$ \\\hline
\end{tabular}
\caption{AND gate truth table where the two inputs are labeled by $A$ and $B$,
the output labeled by $O$, and the $1$ or $0$ entries represent `on' and `off'.
(a)  Truth table describing the device characteristics.  (b)  Numerical
simulation of the atomtronic AND gate normalized to the maximal output current `O'.
Here, the inputs A and B are the respective chemical potentials measured with
respect to the onsite interaction energy ($\mu_A/U$ $\mu_B/U$) and the output
is the atomic current normalized by the maximum output current ($I/I_{max}$).
In practice, the logical `ins' and `outs' are never really `0' or `1' but are
defined by setting threshold values.  Since the maximal output current is
larger than all other output values by at least a factor of $6$, we can
conclude that there exists logic gate behavior in this optical lattice setup.}
\label{Truth}\end{table}

\section{CONCLUSIONS}
In this article, we have derived a general model for treating a specific class
of open quantum systems where the reservoirs act as sources and sinks for
particles moving into and out of the system.  Such a formalism can be used to
study atomic transport across arbitrary multiple potential well configurations.
Here, the formalism was used to show how neutral atoms in custom optical
lattices can exhibit electronic diode, FET, BJT, and AND gate behavior.

Looking forward, we aim to develop more complicated atomtronic devices such as
additional logic elements, flip-flops, and constant current sources by
cascading our current atomtronic components in a manner analogous to the
development more sophisticated electronic devices.  The simulation of the AND
gate is promising since it demonstrates the possibility of cascading atomtronic
components to make more sophisticated devices.

\section{ACKNOWLEDGMENTS} We would like to thank Rajiv Bhat, Brandon Peden,
Brian Seamen, and Jochen Wachter for their helpful discussions.  This work was
supported by AFOSR and NSF.

\bibliographystyle{apsrev}
\bibliography{SysResBib}

\begin{thebibliography}{20}
\expandafter\ifx\csname natexlab\endcsname\relax\def\natexlab#1{#1}\fi
\expandafter\ifx\csname bibnamefont\endcsname\relax
  \def\bibnamefont#1{#1}\fi
\expandafter\ifx\csname bibfnamefont\endcsname\relax
  \def\bibfnamefont#1{#1}\fi
\expandafter\ifx\csname citenamefont\endcsname\relax
  \def\citenamefont#1{#1}\fi
\expandafter\ifx\csname url\endcsname\relax
  \def\url#1{\texttt{#1}}\fi
\expandafter\ifx\csname urlprefix\endcsname\relax\def\urlprefix{URL }\fi
\providecommand{\bibinfo}[2]{#2}
\providecommand{\eprint}[2][]{\url{#2}}

\bibitem[{\citenamefont{Micheli et~al.}(2004)\citenamefont{Micheli, Daley,
  Jaksch, and Zoller}}]{Zoller}
\bibinfo{author}{\bibfnamefont{A.}~\bibnamefont{Micheli}},
  \bibinfo{author}{\bibfnamefont{A.~J.} \bibnamefont{Daley}},
  \bibinfo{author}{\bibfnamefont{D.}~\bibnamefont{Jaksch}}, \bibnamefont{and}
  \bibinfo{author}{\bibfnamefont{P.}~\bibnamefont{Zoller}},
  \bibinfo{journal}{Phys. Rev. Lett.} \textbf{\bibinfo{volume}{93}},
  \bibinfo{pages}{140408} (\bibinfo{year}{2004}).

\bibitem[{\citenamefont{Stickney et~al.}(2007)\citenamefont{Stickney, Anderson,
  and Zozulya}}]{Zozula}
\bibinfo{author}{\bibfnamefont{J.~A.} \bibnamefont{Stickney}},
  \bibinfo{author}{\bibfnamefont{D.~Z.} \bibnamefont{Anderson}},
  \bibnamefont{and} \bibinfo{author}{\bibfnamefont{A.~A.}
  \bibnamefont{Zozulya}}, \bibinfo{journal}{Phys. Rev. A}
  \textbf{\bibinfo{volume}{75}}, \bibinfo{pages}{013608}
  (\bibinfo{year}{2007}).

\bibitem[{\citenamefont{Pepino et~al.}(2009)\citenamefont{Pepino, Cooper,
  Anderson, and Holland}}]{myprl}
\bibinfo{author}{\bibfnamefont{R.~A.} \bibnamefont{Pepino}},
  \bibinfo{author}{\bibfnamefont{J.}~\bibnamefont{Cooper}},
  \bibinfo{author}{\bibfnamefont{D.~Z.} \bibnamefont{Anderson}},
  \bibnamefont{and} \bibinfo{author}{\bibfnamefont{M.~J.}
  \bibnamefont{Holland}}, \bibinfo{journal}{Phys. Rev. Lett.}
  \textbf{\bibinfo{volume}{103}}, \bibinfo{pages}{140405}
  (\bibinfo{year}{2009}).

\bibitem[{\citenamefont{Cardamone et~al.}(2005)\citenamefont{Cardamone,
  Stafford, and Mazumdar}}]{AZG}
\bibinfo{author}{\bibfnamefont{D.~M.} \bibnamefont{Cardamone}},
  \bibinfo{author}{\bibfnamefont{C.~A.} \bibnamefont{Stafford}},
  \bibnamefont{and} \bibinfo{author}{\bibfnamefont{S.}~\bibnamefont{Mazumdar}},
  \bibinfo{journal}{Nano Lett. 6, 2422 (2006)}  (\bibinfo{year}{2005}).

\bibitem[{\citenamefont{Stafford et~al.}(2007)\citenamefont{Stafford,
  Cardamone, and Mazumdar}}]{ArizonaGang}
\bibinfo{author}{\bibfnamefont{C.~A.} \bibnamefont{Stafford}},
  \bibinfo{author}{\bibfnamefont{D.~M.} \bibnamefont{Cardamone}},
  \bibnamefont{and} \bibinfo{author}{\bibfnamefont{S.}~\bibnamefont{Mazumdar}},
  \bibinfo{journal}{Nanotechnology} \textbf{\bibinfo{volume}{18}},
  \bibinfo{pages}{424014} (\bibinfo{year}{2007}).

\bibitem[{\citenamefont{Theis et~al.}(2004)\citenamefont{Theis, Thalhammer,
  Winkler, Hellwig, Ruff, and Grimm}}]{Theis:opticalFR}
\bibinfo{author}{\bibfnamefont{M.}~\bibnamefont{Theis}},
  \bibinfo{author}{\bibfnamefont{G.}~\bibnamefont{Thalhammer}},
  \bibinfo{author}{\bibfnamefont{K.}~\bibnamefont{Winkler}},
  \bibinfo{author}{\bibfnamefont{M.}~\bibnamefont{Hellwig}},
  \bibinfo{author}{\bibfnamefont{G.}~\bibnamefont{Ruff}}, \bibnamefont{and}
  \bibinfo{author}{\bibfnamefont{R.}~\bibnamefont{Grimm}},
  \bibinfo{journal}{Phys. Rev. Lett.} \textbf{\bibinfo{volume}{93}},
  \bibinfo{pages}{123001} (\bibinfo{year}{2004}).

\bibitem[{\citenamefont{Inouye et~al.}(1998)\citenamefont{Inouye, Andrews,
  Stenger, Miesner, Stamper-Kurn, and Ketterle}}]{Inouye:MoleculeFR}
\bibinfo{author}{\bibfnamefont{S.}~\bibnamefont{Inouye}},
  \bibinfo{author}{\bibfnamefont{M.~R.} \bibnamefont{Andrews}},
  \bibinfo{author}{\bibfnamefont{J.}~\bibnamefont{Stenger}},
  \bibinfo{author}{\bibfnamefont{H.-J.} \bibnamefont{Miesner}},
  \bibinfo{author}{\bibfnamefont{D.~M.} \bibnamefont{Stamper-Kurn}},
  \bibnamefont{and} \bibinfo{author}{\bibfnamefont{W.}~\bibnamefont{Ketterle}},
  \bibinfo{journal}{Nature (London)} \textbf{\bibinfo{volume}{392}},
  \bibinfo{pages}{151} (\bibinfo{year}{1998}).

\bibitem[{\citenamefont{Inouye et~al.}(2004)\citenamefont{Inouye, Goldwin,
  Olsen, Ticknor, Bohn, and Jin}}]{Inouye:heteromolecules}
\bibinfo{author}{\bibfnamefont{S.}~\bibnamefont{Inouye}},
  \bibinfo{author}{\bibfnamefont{J.}~\bibnamefont{Goldwin}},
  \bibinfo{author}{\bibfnamefont{M.~L.} \bibnamefont{Olsen}},
  \bibinfo{author}{\bibfnamefont{C.}~\bibnamefont{Ticknor}},
  \bibinfo{author}{\bibfnamefont{J.~L.} \bibnamefont{Bohn}}, \bibnamefont{and}
  \bibinfo{author}{\bibfnamefont{D.~S.} \bibnamefont{Jin}},
  \bibinfo{journal}{Phys. Rev. Lett.} \textbf{\bibinfo{volume}{93}},
  \bibinfo{pages}{183201} (\bibinfo{year}{2004}).

\bibitem[{\citenamefont{Stan et~al.}(2004)\citenamefont{Stan, Zwierlein,
  Schunck, Raupach, and Ketterle}}]{Stan:heteromolecules}
\bibinfo{author}{\bibfnamefont{C.~A.} \bibnamefont{Stan}},
  \bibinfo{author}{\bibfnamefont{M.~W.} \bibnamefont{Zwierlein}},
  \bibinfo{author}{\bibfnamefont{C.~H.} \bibnamefont{Schunck}},
  \bibinfo{author}{\bibfnamefont{S.~M.~F.} \bibnamefont{Raupach}},
  \bibnamefont{and} \bibinfo{author}{\bibfnamefont{W.}~\bibnamefont{Ketterle}},
  \bibinfo{journal}{Phys. Rev. Lett.} \textbf{\bibinfo{volume}{93}},
  \bibinfo{pages}{143001} (\bibinfo{year}{2004}).

\bibitem[{\citenamefont{Bakr et~al.}(2009)\citenamefont{Bakr, Gillen, Peng,
  F{\\"o}lling, and Greiner}}]{GreinNat}
\bibinfo{author}{\bibfnamefont{W.}~\bibnamefont{Bakr}},
  \bibinfo{author}{\bibfnamefont{J.}~\bibnamefont{Gillen}},
  \bibinfo{author}{\bibfnamefont{A.}~\bibnamefont{Peng}},
  \bibinfo{author}{\bibfnamefont{S.}~\bibnamefont{F{\\"o}lling}},
  \bibnamefont{and} \bibinfo{author}{\bibfnamefont{M.}~\bibnamefont{Greiner}},
  \bibinfo{journal}{Nature} \textbf{\bibinfo{volume}{462}}, \bibinfo{pages}{74}
  (\bibinfo{year}{2009}).

\bibitem[{\citenamefont{Mandel et~al.}(2003)\citenamefont{Mandel, Greiner,
  Widera, Rom, {T. W. H\"ansch}, and Bloch}}]{Mandel:entanglementlattices}
\bibinfo{author}{\bibfnamefont{O.}~\bibnamefont{Mandel}},
  \bibinfo{author}{\bibfnamefont{M.}~\bibnamefont{Greiner}},
  \bibinfo{author}{\bibfnamefont{A.}~\bibnamefont{Widera}},
  \bibinfo{author}{\bibfnamefont{T.}~\bibnamefont{Rom}},
  \bibinfo{author}{\bibnamefont{{T. W. H\"ansch}}}, \bibnamefont{and}
  \bibinfo{author}{\bibfnamefont{I.}~\bibnamefont{Bloch}},
  \bibinfo{journal}{Nature (London)} \textbf{\bibinfo{volume}{425}},
  \bibinfo{pages}{937} (\bibinfo{year}{2003}).

\bibitem[{\citenamefont{Strohmaier et~al.}(2007)\citenamefont{Strohmaier,
  Takasu, Gunter, Jordens, Kohl, Moritz, and Esslinger}}]{EsslingerMoritz}
\bibinfo{author}{\bibfnamefont{N.}~\bibnamefont{Strohmaier}},
  \bibinfo{author}{\bibfnamefont{Y.}~\bibnamefont{Takasu}},
  \bibinfo{author}{\bibfnamefont{K.}~\bibnamefont{Gunter}},
  \bibinfo{author}{\bibfnamefont{R.}~\bibnamefont{Jordens}},
  \bibinfo{author}{\bibfnamefont{M.}~\bibnamefont{Kohl}},
  \bibinfo{author}{\bibfnamefont{H.}~\bibnamefont{Moritz}}, \bibnamefont{and}
  \bibinfo{author}{\bibfnamefont{T.}~\bibnamefont{Esslinger}},
  \bibinfo{journal}{Phys. Rev. Lett.} \textbf{\bibinfo{volume}{99}},
  \bibinfo{pages}{220601} (\bibinfo{year}{2007}).

\bibitem[{\citenamefont{Henderson et~al.}(2006)\citenamefont{Henderson, Kelkar,
  Guti{\'e}rrez-Medina, Li, and Raizen}}]{RaizenLi06}
\bibinfo{author}{\bibfnamefont{K.}~\bibnamefont{Henderson}},
  \bibinfo{author}{\bibfnamefont{H.}~\bibnamefont{Kelkar}},
  \bibinfo{author}{\bibfnamefont{B.}~\bibnamefont{Guti{\'e}rrez-Medina}},
  \bibinfo{author}{\bibfnamefont{T.~C.} \bibnamefont{Li}}, \bibnamefont{and}
  \bibinfo{author}{\bibfnamefont{M.~G.} \bibnamefont{Raizen}},
  \bibinfo{journal}{Phys. Rev. Lett.} \textbf{\bibinfo{volume}{96}},
  \bibinfo{pages}{150401} (\bibinfo{year}{2006}).

\bibitem[{\citenamefont{Denschlag et~al.}(2002)\citenamefont{Denschlag,
  Simsarian, Haeffner, McKenzie, Browaeys, Cho, Helmerson, Rolston, and
  Phillips}}]{PhillipsRolston02}
\bibinfo{author}{\bibfnamefont{J.~H.} \bibnamefont{Denschlag}},
  \bibinfo{author}{\bibfnamefont{J.~E.} \bibnamefont{Simsarian}},
  \bibinfo{author}{\bibfnamefont{H.}~\bibnamefont{Haeffner}},
  \bibinfo{author}{\bibfnamefont{C.}~\bibnamefont{McKenzie}},
  \bibinfo{author}{\bibfnamefont{A.}~\bibnamefont{Browaeys}},
  \bibinfo{author}{\bibfnamefont{D.}~\bibnamefont{Cho}},
  \bibinfo{author}{\bibfnamefont{K.}~\bibnamefont{Helmerson}},
  \bibinfo{author}{\bibfnamefont{S.~L.} \bibnamefont{Rolston}},
  \bibnamefont{and} \bibinfo{author}{\bibfnamefont{W.~D.}
  \bibnamefont{Phillips}}, \bibinfo{journal}{J. Phys. B}
  \textbf{\bibinfo{volume}{35}}, \bibinfo{pages}{3095} (\bibinfo{year}{2002}).

\bibitem[{\citenamefont{Jaksch et~al.}(1998)\citenamefont{Jaksch, Bruder,
  Cirac, Gardiner, and Zoller}}]{JakchZoller}
\bibinfo{author}{\bibfnamefont{D.}~\bibnamefont{Jaksch}},
  \bibinfo{author}{\bibfnamefont{C.}~\bibnamefont{Bruder}},
  \bibinfo{author}{\bibfnamefont{J.~I.} \bibnamefont{Cirac}},
  \bibinfo{author}{\bibfnamefont{C.~W.} \bibnamefont{Gardiner}},
  \bibnamefont{and} \bibinfo{author}{\bibfnamefont{P.}~\bibnamefont{Zoller}},
  \bibinfo{journal}{Phys. Rev. Lett.} \textbf{\bibinfo{volume}{81}},
  \bibinfo{pages}{3108} (\bibinfo{year}{1998}).

\bibitem[{\citenamefont{Greiner et~al.}(2002)\citenamefont{Greiner, Mandel,
  Esslinger, H{\"a}nsch, and Bloch}}]{Greiner}
\bibinfo{author}{\bibfnamefont{M.}~\bibnamefont{Greiner}},
  \bibinfo{author}{\bibfnamefont{O.}~\bibnamefont{Mandel}},
  \bibinfo{author}{\bibfnamefont{T.}~\bibnamefont{Esslinger}},
  \bibinfo{author}{\bibfnamefont{T.~W.} \bibnamefont{H{\"a}nsch}},
  \bibnamefont{and} \bibinfo{author}{\bibfnamefont{I.}~\bibnamefont{Bloch}},
  \bibinfo{journal}{Nature (London)} \textbf{\bibinfo{volume}{415}},
  \bibinfo{pages}{39} (\bibinfo{year}{2002}).

\bibitem[{\citenamefont{Cohen-Tannoudji
  et~al.}(1992)\citenamefont{Cohen-Tannoudji, Dupont-Roc, and
  Grynberg}}]{Cohen}
\bibinfo{author}{\bibfnamefont{C.}~\bibnamefont{Cohen-Tannoudji}},
  \bibinfo{author}{\bibfnamefont{J.}~\bibnamefont{Dupont-Roc}},
  \bibnamefont{and} \bibinfo{author}{\bibfnamefont{G.}~\bibnamefont{Grynberg}},
  \emph{\bibinfo{title}{{Atom-Photon interactions}}}
  (\bibinfo{publisher}{Wiley}, \bibinfo{year}{1992}).

\bibitem[{\citenamefont{Meystre and Sargent}(1999)}]{Meistre}
\bibinfo{author}{\bibfnamefont{P.}~\bibnamefont{Meystre}} \bibnamefont{and}
  \bibinfo{author}{\bibfnamefont{M.}~\bibnamefont{Sargent}},
  \emph{\bibinfo{title}{{Elements of Quantum Optics}}}
  (\bibinfo{publisher}{Springer}, \bibinfo{year}{1999}).

\bibitem[{\citenamefont{Walls and Milburn}(1994)}]{Walls}
\bibinfo{author}{\bibfnamefont{D.~F.} \bibnamefont{Walls}} \bibnamefont{and}
  \bibinfo{author}{\bibfnamefont{G.~J.} \bibnamefont{Milburn}},
  \emph{\bibinfo{title}{{Quantum optics}}} (\bibinfo{publisher}{Springer},
  \bibinfo{year}{1994}).

\bibitem[{\citenamefont{Zubarev et~al.}(1997)\citenamefont{Zubarev, Morozov,
  and R{\"o}pke}}]{Zubarev}
\bibinfo{author}{\bibfnamefont{D.~N.} \bibnamefont{Zubarev}},
  \bibinfo{author}{\bibfnamefont{V.}~\bibnamefont{Morozov}}, \bibnamefont{and}
  \bibinfo{author}{\bibfnamefont{G.}~\bibnamefont{R{\"o}pke}},
  \emph{\bibinfo{title}{{Statistical mechanics of nonequilibrium processes. 2.
  Relaxation and hydrodynamic processes}}} (\bibinfo{publisher}{Akad.-Verl.},
  \bibinfo{year}{1997}).

\end{thebibliography}

\end{document}